\documentclass[manuscript]{aastex}

\newcommand{\teff}{$T_{\rm eff}$}
\newcommand{\feh}{\mbox{[Fe/H]}}
\newcommand{\hii}{\mbox{H \sc{ii}}}
\newcommand{\vsini}{\mbox{$v\,\sin\,i$}}
\newcommand{\vmicro}{$\xi$}
\newcommand{\vmacro}{$\Xi$}
\newcommand{\vrad}{\mbox{$v_{\rm rad}$}}

\usepackage{palatcm}
\slugcomment{submitted to the Astronomical Journal}
\shorttitle{Pleiades Metallicity}
\shortauthors{Soderblom et al.}

\begin{document}

\title{The Metallicity of the Pleiades}

\author{David R. Soderblom}
\affil{Space Telescope Science Institute, 3700 San Martin Dr., Baltimore MD 21218}
\email{soderblom@stsci.edu}

\author{Tanmoy Laskar}
\affil{Department of Physics, Cavendish Laboratory, University of Cambridge,\\
J. J. Thomson Avenue, Cambridge CB3 0HE, UK\\
and\\
Space Telescope Science Institute, 3700 San Martin Dr., Baltimore MD 21218}

\author{Jeff A. Valenti}
\affil{Space Telescope Science Institute, 3700 San Martin Dr., Baltimore MD 21218}

\and

\author{John R. Stauffer and Luisa M. Rebull}
\affil{Spitzer Science Center, California Institute of Technology, Pasadena CA 91125}


\begin{abstract}

We have measured the abundances of Fe, Si, Ni, Ti, and Na in 20 Pleiads with  \teff\ values near solar and with low \vsini\ using high-resolution, high signal-to-noise echelle spectra.  We have validated our procedures by also analyzing 10 field stars of a range of temperatures and metallicities that were observed by \citet{Vale05}.  Our result for the Pleiades is [Fe/H] = $+0.03\pm0.02\pm0.05$ (statistical and systematic).  The average of published measurements for the Pleiades is $+0.042\pm0.021$.

\end{abstract}


\section{Introduction}

Star clusters are critical for understanding the properties and processes of our Galaxy, and open clusters particularly so for the disk.  Having a significant number of stars of the same age and composition allows us to test models of stellar evolution, but only if a cluster's composition is well determined.  Clusters also help delineate Galactic gradients in composition, but again that composition must be measured.

Among the Galaxy's open clusters, the Pleiades remains a cornerstone.  It is nearby, making its members accessible to high-resolution spectroscopy.  The precise distance to the Pleiades remains contentious despite efforts to measure the distance in a number of independent ways.  All of those methods are consistent to within the stated uncertainties \citep{Pins98, Nara99, Gate00, Stel01, Maka02, Muna04, Pann04, Zwah04, John05, Sode05, Sout05} with the notable exception of the result from the {\it Hipparcos} mission, although corrections to the {\it Hipparcos} value \citep{vanL07,vanL09} have led to distances nearer those determined by other studies.  Changes in distances and metallicities of the nearby clusters produce similar effects on an H-R diagram \citep{AnTe07}, so having independent measures of those quantities is vital.

The Pleiades is a rich and very well studied cluster.  Quoted turn-off ages include: 50 Myr \citep{Pate78}; 78 Myr \citep{Merm81}; 150 Myr \citep{Mazz89}; 100 Myr \citep{Meyn93}; $\ge 120$ Myr \citep{Vent98}; 120 Myr \citep{Khar05}; 135 Myr (Webda database\footnote{\url{http://www.univie.ac.at/webda}}); and $79\pm52$ Myr \citep{Paun06}.  The lithium depletion boundary (LDB) age is given as 120--130 Myr \citep{Stau98} and $130\pm20$ Myr \citep{Barr04}.  The simple average of these values is approximately $100\pm50$, which in part reflects the large systematic differences in ages determined from the main sequence turn-off as compared to the LDB, a problem that remains unresolved.  In any case, the Pleiades is an exemplar of a Zero-Age Main Sequence cluster for intermediate-mass (about 0.5 to 2.0 $M_\Sun$) stars.

Despite the importance and accessibility of the Pleiades, there have been few determinations of the cluster's metallicity: [Fe/H] = $-0.034\pm0.024$ \citep{Boes90}; $+0.06\pm0.05$ \citep{King00}; $+0.01$ \citep{Barr01}; $+0.06\pm0.23$ \citep{Groe07}; $+0.06\pm0.02$ \citep{Gebr08}.  Some other studies have looked at chemical peculiarities among the A stars, which we do not consider here.  \citet{AnTe07} reanalyzed the results of \citet{Boes90} and eliminated cluster non-members to get [Fe/H] = $+0.03\pm0.02$.  Thus the average measured metallicity for the Pleiades appears to be $\sim10$\% supra-solar.

We were motivated to make a new study of the metallicity of the Pleiades when undertaking an analysis of abundances in the HD 98800 system \citep{Lask09}.  During the same observing run as for HD 98800, we obtained similar spectra of a number of field stars and of Pleiades members, providing a means for us to calibrate our procedures against a well-determined norm and to establish the scatter in looking at a group of stars that we expect to have the same inherent abundances.  Because the Pleiades is young, its late-type dwarfs exhibit significant chromospheric activity \citep{Sode93b}, and that activity may influence the apparent abundances by changing line formation conditions, although we anticipate that any such effect is less in the F dwarfs we have observed than for later spectral types.  HD 98800 is even younger, being a pre-main sequence system, and may be vulnerable to the same effect.  The Pleiades offers a chance to test the influence of activity on metallicity.  Recently \citet[][henceforth VF05]{Vale05} published a comprehensive study of abundances in a large number of nearby F, G, and K dwarfs, using a package called Spectroscopy Made Easy \citep{Vale96}.  The field stars we observed are in the VF05 sample, and our spectra overall are very similar to those used in that study, meaning that we can derive a Pleiades metallicity that is on a consistent basis with the nearby stars and which has been calibrated against the Sun.

\section{Observations and Analysis}

The data analyzed here were obtained over three nights in 2006 December using the Hamilton echelle spectrograph at Lick Observatory \citep{Vogt87}.  Reductions used standard IRAF procedures in the {\it echelle} package.  The projected slit width was 1.2 arcsec, yielding a resolving power of 40,000.

The extraction of abundances from the spectra used the Spectroscopy Made Easy (SME) package of \citet{Vale96}.  SME includes the computation of opacities and elemental ionization fractions in calculating radiative transfer in a stellar atmosphere.  Local thermodynamic equilibrium is assumed through the line-forming regions.  Model atmospsheres are from \citet{Kuru93}.  SME interpolates the atmospheres and uses as input parameters the star's effective temperature (\teff), gravity ($\log g$), metallicity ([M/H]), microturbulent velocity (\vmicro), radial-tangential macroturbulent velocity (\vmacro), projected rotation rate (\vsini), and radial velocity (\vrad).  Any of these parameters may also be calculated from iterative fits to a spectrum.  Note that the [M/H] value is a scaling factor that is applied to solar elemental abundances to generate stellar opacities and it is not a true metallicity in itself.  All velocities are in km s$^{-1}$.

We applied SME to our spectra in a manner essentially identical to that of VF05.  VF05 used SME to determine abundances for Na, Si, Ti, Fe, and Ni in 1,040 F, G, and K stars of the solar neighborhood.  The procedures used are fully described there and will only be summarized here.  We used the same wavelength interval as for the VF05 analyses of Lick spectra: five orders centered near 6100 \AA.  VF05 tuned atomic line data in this interval until they could accurately reproduce the \citet{Wall98} solar spectrum, and they masked lines (particularly Ca features) that could not be consistently reproduced across the \teff\ range of their sample or which did not appear in the line databases.

\subsection{Field Star Validation}

We started by analyzing spectra of 10 F and G field stars that were studied by VF05.  The stars observed and our derived values are listed in Table \ref{tab-field}.   We performed two experiments.  In the first (Case A), we fit the spectrum for \teff, as was done by VF05.  In that case, on average our \teff\ values are 11 K lower than those of VF05, a negligible amount, with a scatter of 76 K.  This scatter seems large since we are fitting spectra nearly identical to those of VF05 and in all the same ways.  Uncertainties of about 100K are taken as typical in determining \teff\ for stars, but that is for comparing independent methods.  In our Case B, we calculated \teff\ from the star's \bv\ value using the relation in \citet{Sode93b}:

$$T_{\rm eff} (K) = 1808(B-V)^{2} -6103(B-V)+8899.$$
\noindent
In that case there is a systematic offset of 67 K in the \teff\ values (in the sense that the we get lower values than did VF05), with an rms scatter of 188 K, an even larger amount.

\begin{figure}
\epsscale{1}
\figurenum{1}
\plotone{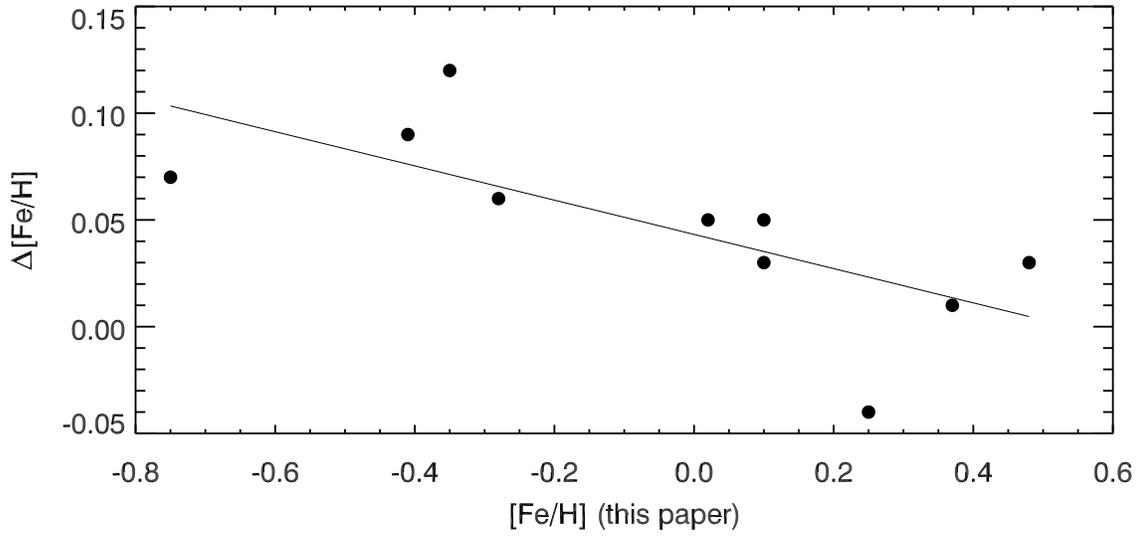}
\caption{\label{fig-diff}A comparison of our measures of [Fe/H] to those in VF05 for the 10 field stars in common.  The ordinate is in the sense of our minus VF05, and the line is a least-squares fit to the points.  The correlation coefficient is 0.71.}
\end{figure}

The mean difference in \feh\ values we derive compared to VF05 is $-0.047\pm0.014$ with rms scatter of 0.044 dex (i.e., our \feh\ values are systematically larger than those of VF05).  For Case B, the mean difference in \feh\ is $-0.059\pm0.012$ with 0.036 dex scatter.  In neither case is the scatter dominated by just one or two stars, and so we conclude that the \feh\ values we determine in this study are systematically about 0.05 dex larger than those in VF05.  Figure \ref{fig-diff} shows the measured differences in \feh\ (our minus VF05) versus our \feh.  There is a correlation, and the mean relation is at a difference of $-0.04$ for a measured \feh\ = +0.08.  In other words, the mean difference is the same as the fitted difference at the [Fe/H] we measure for the Pleiades (see below).  This mean difference between the present study and VF05 is relatively large, but the two datasets were obtained and reduced by different individuals, and so we anticipate that a correction for scattered light, for instance, could lead to the effect seen.

\subsection{Pleiades Analysis}

We analyzed spectra of 20 F, G, and K dwarf members of the Pleiades in the same manner as the above field stars, and the results are in Table \ref{tab-Pleiades}.  We also obtained spectra of three binaries in the Pleiades that have components of nearly equal mass (\hii\ 739, 1117, and 1726) but have not included them here.  The $(B - V)_0$ (dereddened) colors are from \citet{Sode93b}, except for star \hii\ 1200, for which the color was estimated from the spectral type because it has anomalous reddening.  We first used SME to fit for the radial velocity offset of each Pleiad, fixing all the other parameters at nominal values. Having determined \vrad, we then fit for \teff\ and the abundances of the five elements (Na, Si, Ti, Ni and Fe), as well as for \vsini, by first shifting the observed spectrum to the laboratory frame.

\section{Results}

\subsection{The metallicity of the Pleiades}

The mean Pleiades \feh\ value we measure is $+0.075\pm0.011$.  There are a few high values of \feh, but the mean is not significantly skewed by these because the median value is 0.07.  After applying the correction factor noted above, our mean Pleiades \feh\ is $+0.03\pm0.02$, where the uncertainty is the statistical value that also takes into account the uncertainty in the correction.  A minimum estimate for the systematic error is 0.05, the scatter in our values of \feh\ for the field stars compared to VF05.

This mean \feh\ is consistent with nearly all the previous determinations (Table \ref{tab-feh}) that have been published.  The most frequently cited measurement is that of \citet{Boes90}: $-0.034\pm0.024$, in part because data of very high quality were used and because the sample consisted of F stars, the Fs being nearly enough solar-like to be able to work differentially yet not as likely to be influenced by chromospheric activity as the G and K dwarfs in the Pleiades.  However, \citet{AnTe07} reexamined what \citet{Boes90} had published, deleted some non-members and took account of redundant information to come up with a mean \feh\ from the same observations of $+0.03\pm0.02$, a significant readjustment.



\citet{Pins98} investigated the effect of a non-solar metallicity on the color-magnitude diagram (CMD) of the Pleiades as a means of explaining the small distance modulus found by {\em Hipparcos}.  A high metallicity for the Pleiades tends to make the cluster's distance discrepancy worse.  Matching the {\em Hipparcos} distance would require \feh\ values of $-0.25$ to $-0.45$, depending on which color one uses, $(B-V)$ or $(V-I)$, because lowering the metallicity lowers the luminosity of stars at a given color.  As discussed in the introduction, independent determinations of the Pleiades distance have not been able to verify the {\em Hipparcos} result and instead have found distances in agreement with traditional main sequence fitting.

\section{Conclusion}

The net result is that we determine [Fe/H] = $+0.03\pm0.02\pm0.05$ for the Pleiades, where the first uncertainty is purely statistical and the second is the systematic.  The measured metallicity is lowered 0.05 dex by applying a correction for the mean difference that we derive for stars in common with VF05.  This mean offset is $-0.047\pm0.044$, consistent with zero, but there is a trend of difference with [Fe/H] (Fig. \ref{fig-diff}) which leads to the same correction factor.  Our determination for the \feh\ of the Pleiades is on the same overall scale as the extensive work on nearby solar-type stars by \citet{Vale05}, which means that it is calibrated against the Sun.

The average of the measured \feh\ values for the Pleiades in Table \ref{tab-feh} is $+0.042\pm0.021$; this was calculated without including \citet{Boes90}, which was corrected by \citet{AnTe07}, nor \citet{Tayl08}, which was not an independent measurement.  The quoted uncertainty is the statistical one and does not include any systematic error.  Thus a number of independent studies lead to a consistent result, namely that the Pleiades is about 10\% richer in heavy elements than the Sun.

\begin{deluxetable}{rcc cc cc cc cc cc cc}
\tabletypesize{\tiny}
\tablewidth{0pt}
\tablecaption{Derived Abundances of Field Stars\label{tab-field}}
\tablehead{
\colhead{HD} & $(B - V)$ & Sp. & \multicolumn{2}{c}{$T_{\rm eff}$} 
 & \multicolumn{2}{c}{[Na/H]} & \multicolumn{2}{c}{[Si/H]} 
  & \multicolumn{2}{c}{[Ti/H]} & \multicolumn{2}{c}{[Ni/H]} & \multicolumn{2}{c}{[Fe/H]}\\
\cline{4-5}  \cline{6-7}  \cline{8-9}  \cline{10-11}  \cline{12-13}  \cline{14-15} 
 \colhead{} & \colhead{} & \colhead{} & \colhead{VF05\tablenotemark{a}} & \colhead{SLVS\tablenotemark{b}} & 
 \colhead{VF05} & \colhead{SLVS} & \colhead{VF05} & \colhead{SLVS} & 
  \colhead{VF05} & \colhead{SLVS} & \colhead{VF05} & \colhead{SLVS} & \colhead{VF05} & \colhead{SLVS} \\
  \cline{1-15}
  \multicolumn{15}{c}{a) $T_{\rm eff}$ from spectrum fitting}
   }
   \startdata
  1832 & 0.63 & F8V & 5731 & 5793 &  $-0.07$ & $+0.01$ & $-0.04$ & $+0.02$ & $-0.04$ & $-0.04$ & $-0.06$ & $-0.07$ & $-0.03$ & $+0.02$\\
  4203 & 0.76 & G5V & 5702 & 5777 &  $+0.50$ & $+0.61$ & $+0.40$ & $+0.45$ & $+0.40$ & $+0.52$ & $+0.47$ & $+0.50$ & $+0.45$ & $+0.48$\\
  9472 & 0.62 & G0V & 5867 & 5868 &  $-0.10$ & $-0.03$ & $-0.02$ & $+0.04$ & $+0.05$ & $+0.03$ & $-0.06$ & $-0.06$ & $+0.05$ & $+0.10$\\
 12414 & 0.44 & F2V & 6158 & 6030 &  $-0.40$ & $-0.26$ & $-0.29$ & $-0.30$ & $-0.25$ & $-0.35$ & $-0.44$ & $-0.46$ & $-0.34$ & $-0.28$\\
 12661 & 0.72 & K0V & 5743 & 5686 &  $+0.51$ & $+0.61$ & $+0.35$ & $+0.36$ & $+0.32$ & $+0.25$ & $+0.42$ & $+0.41$ & $+0.36$ & $+0.37$\\
45350 & 0.74 & G5V & 5616 & 5487 &  $+0.27$ & $+0.25$ & $+0.27$ & $+0.26$ & $+0.24$ & $+0.13$ & $+0.26$ & $+0.23$ & $+0.29$ & $+0.25$\\
45391 & 0.60 & G0V & 5624 & 5673 &  $-0.41$ & $-0.36$ & $-0.40$ & $-0.34$ & $-0.41$ & $-0.50$ & $-0.51$ & $-0.43$ & $-0.50$ & $-0.41$\\
63433 & 0.64 & G5IV & 5742 & 5702 &  $-0.09$ & $-0.08$ & $-0.01$ & $+0.02$ & $+0.07$ & $+0.04$ & $-0.05$ & $-0.04$ & $+0.07$ & $+0.10$\\
102158 & 0.61 & G2V & 5725 & 5735 &  $-0.35$ & $-0.24$ & $-0.21$ & $-0.14$ & $-0.13$ & $-0.11$ & $-0.41$ & $-0.39$ & $-0.47$ & $-0.35$\\
104800 & 0.59 & G0V & 5626 & 5677 &  $-0.81$ & $-0.61$ & $-0.54$ & $-0.50$ & $-0.51$ & $-0.50$ & $-0.83$ & $-0.74$ & $-0.82$ & $-0.75$\\
  \cline{1-15}
  \multicolumn{3}{r}{avg. $\Delta$} & \multicolumn{2}{c}{$+11$} &  \multicolumn{2}{c}{} 
  & \multicolumn{2}{c}{} & \multicolumn{2}{c}{} & \multicolumn{2}{c}{} & \multicolumn{2}{c}{$-0.047$}\\
    \multicolumn{3}{r}{$\sigma$} & \multicolumn{2}{c}{$\pm76$} &  \multicolumn{2}{c}{} 
  & \multicolumn{2}{c}{} & \multicolumn{2}{c}{} & \multicolumn{2}{c}{} & \multicolumn{2}{c}{$\pm0.044$}\\
    \multicolumn{3}{r}{$\sigma/\surd N$} & \multicolumn{2}{c}{$\pm24$}  & \multicolumn{2}{c}{} 
  & \multicolumn{2}{c}{} & \multicolumn{2}{c}{} & \multicolumn{2}{c}{} & \multicolumn{2}{c}{$\pm0.014$}\\
    \cline{1-15}
    \multicolumn{15}{c}{b) $T_{\rm eff}$ from $(B - V)$}\\
    \cline{1-15}
    1832 & 0.63 & F8V &   5731 & 5733 &    $-0.07$ & $0.00$ & $-0.04$ & $+0.01$ & $-0.04$ & $-0.11$ & $-0.06$ & $-0.08$ & $-0.03$ & $+0.01$\\
    4203 & 0.76 & G5V &  5702 & 5315 &  $+0.50$ & $+0.62$ & $+0.40$ & $+0.44$ & $+0.40$ & $+0.50$ & $+0.47$ & $+0.47$ & $+0.45$ & $+0.49$\\
    9472 & 0.62 & G0V &  5867 & 5810 &   $-0.10$ &  $+0.06$ & $-0.02$ & $0.00$ & $+0.05$ & $+0.08$ & $-0.06$ & $-0.16$ & $+0.05$ & $+0.17$\\
  12414 & 0.44 & F2V &  6158 & 6299 &   $-0.40$ &  $-0.19$ & $-0.29$ & $-0.26$ & $-0.25$ & $-0.27$ & $-0.44$ & $-0.38$ & $-0.34$ & $-0.23$\\
  12661 & 0.72 & K0V &  5743 & 5442 &  $+0.51$ & $+0.65$ & $+0.35$ & $+0.35$ & $+0.32$ & $+0.27$ & $+0.42$ & $+0.41$ & $+0.36$ & $+0.38$\\
  45350 & 0.74 & G5V &  5616 & 5372 &  $+0.27$ & $+0.32$ & $+0.27$ & $+0.30$ & $+0.24$ & $+0.25$ & $+0.26$ & $+0.28$ & $+0.29$ & $+0.33$\\
  45391 & 0.60 & G0V &  5624 & 5584 &   $-0.41$ &  $-0.40$ & $-0.40$ & $-0.36$ & $-0.41$ & $-0.56$ & $-0.51$ & $-0.49$ & $-0.50$ & $-0.47$\\
  63433 & 0.64 & G5IV & 5742 & 5733 & $-0.09$ &  $-0.03$ & $-0.01$ & $+0.04$ & $+0.07$ & $+0.09$ & $-0.05$ & $-0.02$ & $+0.07$ & $+0.12$\\
102158 & 0.61 & G2V &  5725 & 5771 &   $-0.35$ &  $-0.27$ & $-0.21$ & $-0.14$ & $-0.13$ & $-0.08$ & $-0.41$ & $-0.43$ & $-0.47$ & $-0.37$\\
104800 & 0.59 & G0V &  5626 & 5810 &   $-0.81$ &  $-0.63$ & $-0.54$ & $-0.51$ & $-0.51$ & $-0.54$ & $-0.83$ & $-0.77$ & $-0.82$ & $-0.78$\\
  \cline{1-15}
  \multicolumn{3}{r}{avg. $\Delta$} & \multicolumn{2}{c}{$+67$} &  \multicolumn{2}{c}{} 
  & \multicolumn{2}{c}{} & \multicolumn{2}{c}{} & \multicolumn{2}{c}{} & \multicolumn{2}{c}{$-0.059$}\\
    \multicolumn{3}{r}{$\sigma$} & \multicolumn{2}{c}{$\pm188$} &  \multicolumn{2}{c}{} 
  & \multicolumn{2}{c}{} & \multicolumn{2}{c}{} & \multicolumn{2}{c}{} & \multicolumn{2}{c}{$\pm0.036$}\\
    \multicolumn{3}{r}{$\sigma/\surd N$} & \multicolumn{2}{c}{$\pm59$}  & \multicolumn{2}{c}{} 
  & \multicolumn{2}{c}{} & \multicolumn{2}{c}{} & \multicolumn{2}{c}{} & \multicolumn{2}{c}{$\pm0.012$}\\
\enddata
\tablenotetext{a}{\citet{Vale05}}
\tablenotetext{b}{This paper.}

\end{deluxetable}

\begin{deluxetable}{rc cr rr rr r}
\tabletypesize{\scriptsize}
\tablewidth{0pt}
\tablecaption{Derived Abundances of Pleiades Members\label{tab-Pleiades}}
\tablehead{
\colhead{H II} & $(B - V)_0$ &  \colhead{$T_{\rm eff}$} & \colhead{$v \sin i$}
 &  \colhead{[Na/H]} & \colhead{[Si/H]} 
  & \colhead{[Ti/H]} & \colhead{[Ni/H]} & \colhead{[Fe/H]}
   }
   \startdata
  120	&	0.666	&	5675	&	11.9	&	$-0.03$	&	$-0.05$	&	$0.00$	&	$-0.20$	&	$+0.07$	\\
233	&	0.493	&	6409	&	14.3	&	$-0.07$	&	$-0.04$	&	$-0.11$	&	$-0.03$	&	$+0.02$	\\
489	&	0.594	&	5921	&	17.2	&	$-0.21$	&	$-0.10$	&	$-0.01$	&	$-0.16$	&	$+0.01$	\\
514	&	0.657	&	5811	&	11.5	&	$-0.10$	&	$+0.06$	&	$-0.09$	&	$-0.08$	&	$+0.05$	\\
761	&	0.652	&	5755	&	12.2	&	$+0.19$	&	$-0.06$	&	$+0.01$	&	$-0.13$	&	$+0.06$	\\
923	&	0.579	&	5821	&	18.6	&	$-0.19$	&	$-0.01$	&	$+0.14$	&	$-0.11$	&	$+0.19$	\\
996	&	0.603	&	5874	&	11.6	&	$-0.09$	&	$-0.04$	&	$-0.01$	&	$-0.14$	&	$+0.08$	\\
1015	&	0.609	&	5848	&	11.0	&	$-0.07$	&	$-0.05$	&	$-0.12$	&	$-0.13$	&	$+0.04$	\\
1182	&	0.597	&	5876	&	16.7	&	$-0.05$	&	$-0.03$	&	$+0.09$	&	$-0.06$	&	$+0.10$	\\
1200	&	0.45	  	&	6535	&	14.2	&	$+0.21$	&	$0.00$	&	$-0.11$	&	$+0.01$	&	$+0.04$	\\
1207	&	0.591	&	5892	&	5.3	&	$-0.16$	&	$+0.02$	&	$0.00$	&	$-0.10$	&	$+0.08$	\\
1215	&	0.599	&	5867	&	6.4	&	$-0.11$	&	$-0.03$	&	$-0.08$	&	$-0.11$	&	$+0.06$	\\
1514	&	0.609	&	5817	&	15.2	&	$+0.02$	&	$+0.08$	&	$+0.24$	&	$-0.09$	&	$+0.18$	\\
1794	&	0.589	&	5899	&	14.8	&	$-0.12$	&	$+0.04$	&	$+0.06$	&	$-0.02$	&	$+0.09$	\\
1797	&	0.515	&	6060	&	21.1	&	$-0.24$	&	$0.00$	&	$0.00$	&	$-0.15$	&	$0.00$	\\
1856	&	0.514	&	6167	&	16.1	&	$+0.11$	&	$+0.08$	&	$+0.11$	&	$+0.02$	&	$+0.13$	\\
1924	&	0.568	&	5879	&	14.5	&	$-0.21$	&	$-0.04$	&	$-0.08$	&	$-0.14$	&	$+0.03$	\\
2172	&	0.582	&	5949	&	11.5	&	$-0.17$	&	$-0.02$	&	$+0.02$	&	$-0.10$	&	$+0.05$	\\
2506	&	0.552	&	6058	&	14.2	&	$-0.01$	&	$+0.07$	&	$+0.19$	&	$-0.07$	&	$+0.13$	\\
3179	&	0.529	&	6203	&	5.3	&	$+0.04$	&	$+0.02$	&	$-0.08$	&	$+0.01$	&	$+0.08$	\\
    \cline{1-9}
  \multicolumn{4}{r}{avg.} &  \multicolumn{4}{c}{} & \multicolumn{1}{r}{$+0.075$}\\ 
    \multicolumn{4}{r}{$\sigma$} &  \multicolumn{4}{c}{} & \multicolumn{1}{r}{$\pm0.051$}\\ 
  \multicolumn{4}{r}{$\sigma/\surd N$} &  \multicolumn{4}{c}{} & \multicolumn{1}{r}{$\pm0.011$}\\ 

\enddata
\end{deluxetable}

\begin{deluxetable}{lc}
\tablewidth{0pt}
\tablecaption{Determinations of [Fe/H] for the Pleiades\label{tab-feh}}
\tablehead{
\colhead{Source} & \colhead{[Fe/H]}}

\startdata
\citet{Boes90} & $-0.034\pm0.024$\\
\citet{Barr01} & $+0.01$\\
\citet{King00} & $+0.06\pm0.05$\\
\citet{Groe07} & $+0.06\pm0.23$\\
\citet{AnTe07} & $+0.03\pm0.02$\tablenotemark{a}\\
\citet{Tayl08} & $-0.039\pm0.014$\tablenotemark{b}\\
\citet{Gebr08} & $+0.06\pm0.02$\\
this study & $+0.03\pm0.04$\\
\enddata
\tablenotetext{a}{Results of \citet{Boes90} recalculated after deleting non-members.}
\tablenotetext{b}{Reanalysis of literature values, not a separate determination.}
\end{deluxetable}

\end{document}